\documentclass{nature}

\usepackage{graphicx}
\usepackage{times,amsmath}
\usepackage{epsfig}
\usepackage{color}
\usepackage{graphicx}
\usepackage{dcolumn}
\usepackage{bm}
\usepackage{bookmark}
\usepackage{tabularx}
\usepackage{hyperref}
\usepackage{multirow}
\usepackage{array,mathtools,amssymb,booktabs,makecell}
\hypersetup{colorlinks=true, citecolor=blue, filecolor=blue, linkcolor=blue, urlcolor=blue}

\usepackage{float}

\usepackage{upgreek}
\makeatletter
\let\saved@includegraphics\includegraphics
\AtBeginDocument{\let\includegraphics\saved@includegraphics}

\makeatother

\title{Impacts of \textit{\textbf{f}}-\textit{\textbf{d}} Kondo cloud on superconductivity of nickelates}

\author{Byungkyun Kang$^{1\ast}$, Hyunsoo Kim$^{2}$, Qiang Zhu$^{1}$ \& Chul Hong Park$^{3}$}
 
\begin{document}

\maketitle

\begin{affiliations}
\item Department of Physics and Astronomy, University of Nevada, Las Vegas, Nevada 89154, USA
\item Department of Physics, Missouri University of Science and Technology, Rolla, MO 65409, USA
\item Quantum Matter Core-Facility and Research Center of Dielectric and Advanced Matter Physics, Pusan National University, Busan 46240, Republic of Korea

$^{\ast}$email: byungkyun.kang@unlv.edu

\end{affiliations}


\begin{abstract}
The discovery of superconducting nickelates reignited hope for elucidating the high-$T_{\textrm{c}}$ superconductivity mechanism in the isostructural cuprates.
While in the cuprates, the superconducting gap opens up on a single-band of the quasi-2D Fermi surface, the nickelates are known to have 3D nature of electronic structure with multi-band.
This raises a serious question about the role of 2D nature for the high-$T_{\textrm{c}}$ superconductivity.
Here, employing dynamical mean field theory combined with GW method, we found the Kondo effect driven by the strong correlation of Nd-4$f$ and Ni-3$d$ electrons emerging at low temperature.
The Kondo effect modifies the topology of the Fermi surface leading to 3D multi-band nature. 
Remarkably, the Kondo effect is easily destroyed by lattice modulation, leading to the quasi-2D nature.
Our findings clearly explain the inconsistent occurrence of superconductivity and distinct electrical resistivity behavior between NdNiO$_{2}$ bulk and films.

\end{abstract}

The mechanism of superconductivity through electronic channels has led to remarkable achievements of high-$T_{\textrm{c}}$ superconductivity. The copper- and iron-based superconductors are among the best examples where the intricate interaction between electron and low-energy bosonic excitation overcomes the Coulomb repulsion to form the Cooper pairs\cite{Chubukov2012,Taillefer2010}.
Such a type of superconductivity requires strong electronic correlations and the sign-change in the superconducting energy gap, i.e. nodal superconductivity\cite{Taillefer2010} or sign-changing multi s-wave\cite{Chubukov2012}.
An understanding of the electronic structure is of utmost importance to clarify the superconducting mechanism.

A common thread in the electronic structures of these two families of copper- and iron-based superconductors is quasi-2D nature which provides a strong restriction on the possible pairing symmetries\cite{bollinger_sst2016,qimiao_natrev2016}.
Nevertheless, the discovery of nickelate superconductors\cite{danfeng_nature2019} poses intriguing questions about the quasi-2D essence. 
In comparison with cuprates, whose Fermi surface displays the single-band of the Cu-$d$ and O-$p$ orbital characters\cite{patrick_rmp2006,warren_rmp1989}, the nickelate's Fermi surface is suggested to have multi-band: Ni-3$d_{x^{2}-y^{2}}$ band and another mixed band including rare-earth-5$d$\cite{botana_prx2020,jonathan_prx2020,hirofumi_prl2020,francesco_prx2020,emily_prx2021,liu_prb2021}, where rare-earth-4$f$ is absent.
Despite that the nickelates have layered crystal structure, the measured isotropic upper critical field reveals that the electronic structure does not exhibit the quasi-2D behavior \cite{bai_natuer2021}.
On the contrary, Sun et al. showed the superconductivity in nickelate is anisotropic and it retains the quasi-2D picture \cite{wenjie_arxiv2022}.
A theoretical study, treating the Nd-4$f$ as non-correlated electrons, suggested that chemical doping can drive the multi-band nickelates into one-band Hubbard system\cite{motoharu_npjQM}, in agreement with monotonic increase of temperature at which $R_{\textrm{H}}$ changes its sign with increasing doping level in Nd$_{1-x}$Sr$_{x}$NiO$_{2}$/SrTiO$_{3}$\cite{shengwei_prl2020}. Zeng et al., however, showed that the negligible doping-dependent temperature of $R_{\textrm{H}}$-sign-change for $x$ $\geq$ 0.23 in La$_{1-x}$Ca$_{x}$NiO$_{2}$/SrTiO$_{3}$ \cite{zeng_sciadv2022}. Zeng et al. also pointed out possible impact of lattice constant on the superconductivity in a variety of doped bulk and thin film nickelates\cite{zeng_sciadv2022}. 
These controversial reports call for a rigorous theoretical investigation of the inter-atomic interactions responsible for the 3D nature. 

The discoveries of superconductivity in various rare-earth elements (La\cite{zeng_sciadv2022,wenjie_arxiv2022}, Pr\cite{motoki_nanolett2020}, Nd\cite{danfeng_nature2019}) nickelates suggest that superconductivity may not be influenced by the presence of 4$f$ electrons in the rare-earth layer.
However, understanding the role of Nd-4$f$ in NdNiO$_{2}$, particularly at the vicinity of the Fermi level is an important and controversial subject from a theoretical point of view. 
By the dynamical mean field theory combined with density functional theory (DFT+DMFT), Liu et al. argued that the Nd-4$f$ electrons impact the electronic states far from the Fermi level, leading to unaffected electronic states close to the Fermi level\cite{liu_prb2021}. 
By DFT study, Choi et al. showed that an intra-atomic exchange coupling between the Nd-4$f$ spin and the Nd-5$d$ state in the magnetic ordered system can affect the state at the Fermi-level, and suggested anti-Kondo coupling of the local moment to the conduction bands\cite{choi_prb2020}.
Non-negligible hybridization between Nd-4$f$ and Ni-3$d$ near the Fermi level was found in spin polarized simulations\cite{xiao_ndnio2,jiang_prb2019}. 
Nevertheless, there is no sign of a long-range magnetic ordering down to 2.0 K for NdNiO$_{2}$\cite{hayward_sss2003} and LaNiO$_{2}$\cite{ortiz_prr2022}.

The $f$-electron Kondo effect is proposed to be the key to the understanding of  heavy-fermion superconductivity \cite{Stewart1984}, but its role has been overlooked in high-$T_{\textrm{c}}$ superconductivity. 
There has been no report on the Kondo effect involving Nd-4$f$ electrons in nickelate, to the best of our knowledge. Only the Kondo effect arising from Ni-3$d_{x^{2}-y^{2}}$ electrons has been suggested \cite{frank_prx2020,yi_fronphy2020,danfeng_nature2019,yuhao_comphy2020}, 

In this work, we investigate the temperature dependence and the lattice modulation effect on the electronic structure of nickelates in paramagnetic normal state through $ab$-$initio$ many-body approach, dynamical mean field theory (DMFT) combined with  linearized quasi-particle self-consistent GW (LQSGW) method. 
We identified the Kondo screening of the localized Nd-4$f$ by the itinerant Ni-3$d$ and Nd-5$d$ electrons in NdNiO$_{2}$, emerging at low temperatures. The Kondo effect apparently leads to the multi-band with the 3D nature at the Fermi level.
We also found the strong hybridized peaks of rare-earth-5$d$ and Ni-3$d$ around -1 eV in NdNiO$_{2}$ and LaNiO$_{2}$, which decreases the density of states at the Fermi level at low temperature.
However, both effects are tunable by lattice modulation. By increasing the interlayer distance, the Fermi surface can be transformed from 3D to 2D and the DOS from Ni-3$d$ at the Fermi level increases.
Lattice modulation should thus affect superconductivity in nickelates.
\begin{figure*}[ht]
\centering
\includegraphics[width=0.9 \textwidth]{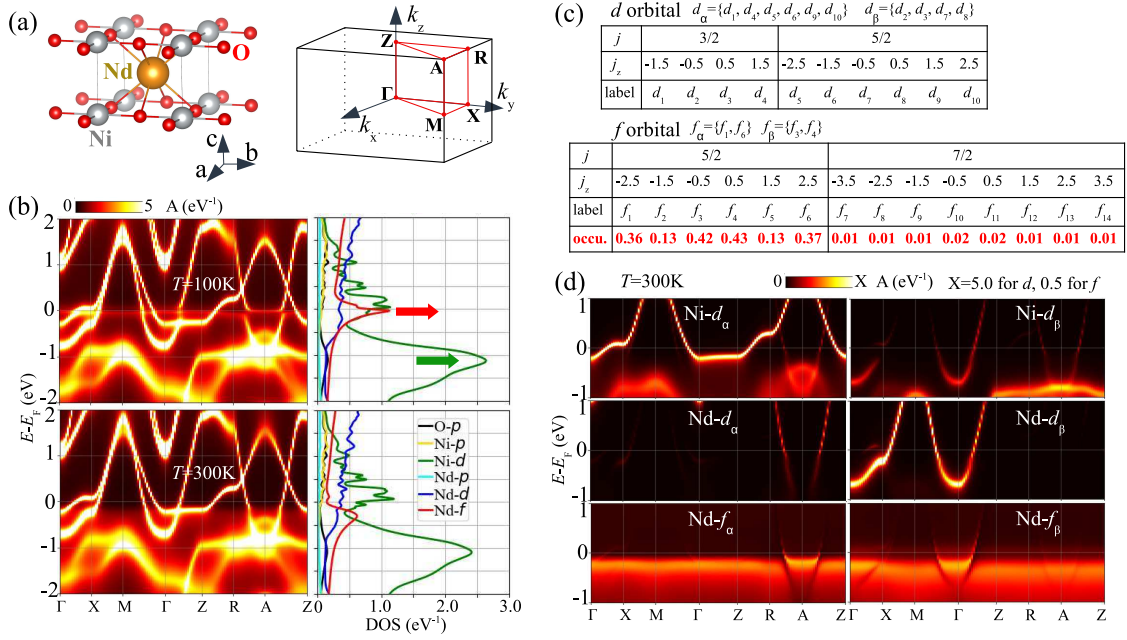}
\caption{\label{Fig_1} \textbf{Kondo effect in NdNiO$_{2}$.} (a) Layered crystal structure of NdNiO$_{2}$ and the high symmetry lines in the first Brillouin Zone. (b) Spectral functions and density of states at 100 K and 300 K. The red and green arrows indicate enhanced peaks at 100 K. (c) $d$ and $f$ orbitals are labelled and grouped for convenience in this work. Electron occupation of Nd-4$f$ are shown. (d) Orbitals projected spectral functions at 300 K.}
\end{figure*}

\section*{Results}
\begin{figure*}[ht]
\centering
\includegraphics[width=0.98 \textwidth]{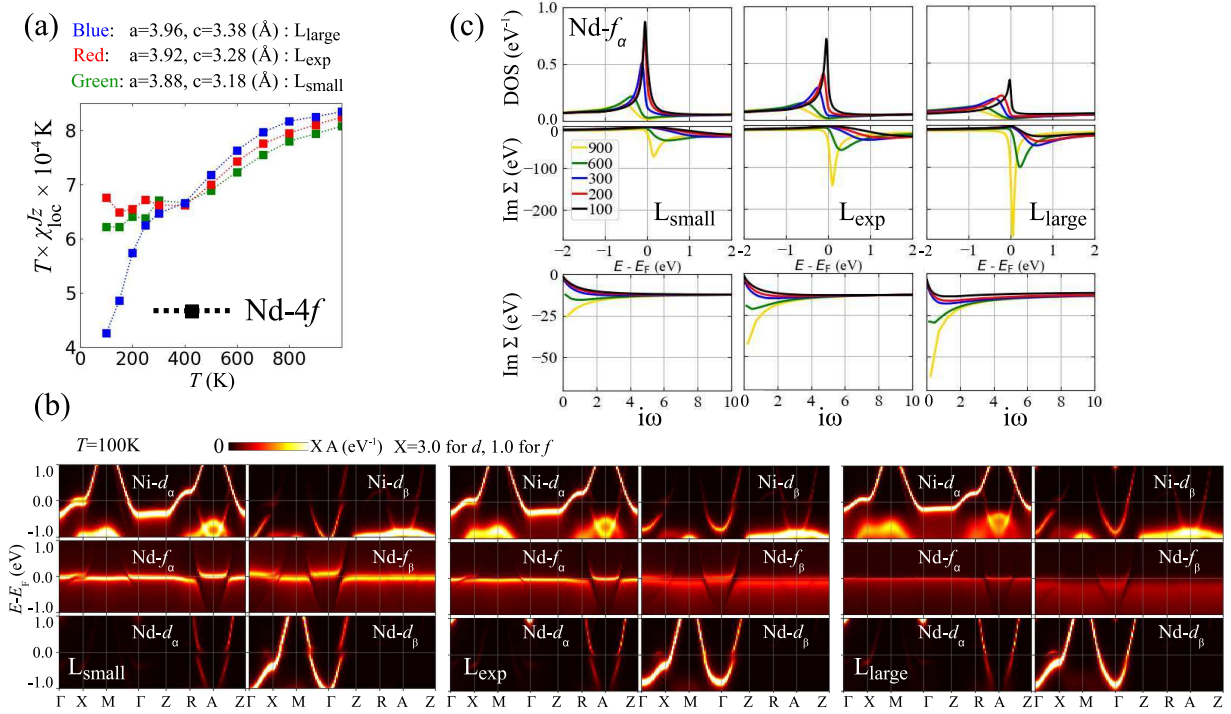}
\caption{\label{Fig_2} \textbf{Kondo effect under lattice modulation.} (a) $T \times \chi_{\mathrm{loc}}^{J_Z}$ of Nd-4$f$ as a function of temperature in three lattices (L$_\mathrm{small}$, L$_\mathrm{exp}$ and L$_\mathrm{large}$). (b) Orbitals projected spectral functions in the three lattices  at 100 K. (c) Density of states, the imaginary part of the self-energy ($\Sigma $) on the imaginary and real frequency axis of Nd-$f_{\alpha}$ are presented. The temperature unit is K.}
\end{figure*}

\textbf{Kondo effect in NdNiO$_{2}$.} We found the Kondo effect driven by the strong correlation of  Nd-4$f$ and Ni-3$d$ electrons. Figure~\ref{Fig_1}(a) shows crystal structure of tetragonal phase (P4/mmm) NdNiO$_{2}$, and the high symmetry lines in the first Brillouin Zone.
Rare-earth Nd atoms make bridges between the NiO$_{2}$-layers.
Figure~\ref{Fig_1}(b) shows spectral functions and density of states (DOS) of NdNiO$_{2}$ at $T=$ 100 and 300 K. 
There are two prominent differences between electronic structures at these two temperatures.
First, the energies of Nd-$f$-driven states increase significantly as temperature decreases.
The DOS of Nd-4$f$ (red line) became sharp and the flat Nd-4$f$ bands appeared at the Fermi level ($E_{\textrm{F}}$) in the spectra function at 100 K.
The flat Nd-4$f$ bands are hybridized with the mixed conduction band of Ni-3$d$ and Nd-5$d$ and give rise to a kink-like band structure at the Fermi level along $\Gamma$-Z and R-A-Z high symmetry lines. That is a hallmark of the Kondo screening\cite{byung_ute2}. 
The table in Figure~\ref{Fig_1}(c) shows significant occupancy of Nd-4$f$ in $j$=5$/$2 states.
The calculated total occupation in Nd-4$f$ orbitals is 1.9 for both temperatures, indicating a large local magnetic moment of Nd-4$f$ is screened by conduction electrons at low temperatures.
This result is reminiscent of heavy fermion superconductor UTe$_{2}$, where flat U-5$f$ bands with occupancy of 2.27 lead to orbital selective Kondo effect\cite{byung_ute2}.
For convenience in this work, $d$ orbitals are labelled, and grouped based on the main orbital character at the vicinity of the Fermi level (-0.5 $<$ $E$-$E_{\textrm{F}}$ $<$ 0.5 eV) as shown in Fig.~\ref{Fig_1}(c).
As shown in Fig.~\ref{Fig_1}(d), Ni-$d_{\alpha}$ is hybridized with Nd-$d_{\alpha}$ along the R-A-Z high symmetry line. Ni-$d_{\beta}$ is hybridized with Nd-$d_{\beta}$ along the $\Gamma$-X and M-R-Z high symmetry lines.
We also group Nd-4$f$ orbitals into $f_{\alpha}$ and $f_{\beta}$ based on occupation. 
At $T=$ 300 K, Nd-$f_{\alpha}$ is hybridized with Nd-$d_{\alpha}$, Ni-$d_{\alpha}$, and Ni-$d_{\beta}$, whereas Nd-$f_{\beta}$ is hybridized with Nd-$d_{\beta}$ and Ni-$d_{\beta}$ in the vicinity of the Fermi level.
At $T=$ 100 K, three Kondo scatterings appeared with momentum-dependent (see Fig.~\ref{Fig_1}(b) and Fig.~\ref{Fig_2}(b) ): i) Nd-$f_{\alpha}$ with Nd-$d_{\alpha}$, Ni-$d_{\alpha}$ and Ni-$d_{\beta}$ along R-A-Z high symmetry line, ii) Nd-$f_{\beta}$ with Nd-$d_{\beta}$ and Ni-$d_{\beta}$ at the X-point and along the M-$\Gamma$-Z high symmetry line, iii) Nd-$f_{\alpha}$ with Ni-$d_{\alpha}$ at the X point and near the $\Gamma$ point.
This indicates that not only Nd-5$d$ but also Ni-3$d$ in NiO$_{2}$ layer are involved to conduction electrons, which screen local spin momentum of Nd-4$f$.
Therefore, the inter-atomic Kondo effect changes the topology of the Fermi surface and leads to the 3D nature of electronic structure at the Fermi surface of NdNiO$_{2}$ at low temperatures.
Another important aspect is that the DOS of Ni-3$d$-driven states (green line) around -1 eV becomes slightly higher with lowering temperature. 
This describes the hybridization between Ni-3$d$ and Nd-5$d$. The details will be demonstrated in next section. 
 
We examined the effect of the lattice modulation on the Nd-4$f$-driven Kondo screening in NdNiO$_{2}$.
The Kondo effect, which manifests an inter-atomic interaction between Nd-4$f$ and the itinerant  Ni-3$d$ electrons, should be sensitive to the distance between Ni and Nd, {\it i.e.}, the lattice modulation, 
owing to localized characters of strongly correlated Nd-4$f$ and Ni-3$d$ electrons.
We use two artificial lattices, namely L$_{\mathrm{large}}$ and L$_\mathrm{small}$, in addition to the experimental lattice constants L$_{\mathrm{exp}}$ ($a$ = 3.92 $\textrm{\AA}$, $c$ = 3.28 $\textrm{\AA}$) of bulk NdNiO$_{2}$\cite{hayward_sss2003}.
For larger lattice L$_{\mathrm{large}}$ ($a$ = 3.96 $\textrm{\AA}$, $c$ = 3.38 $\textrm{\AA}$), we adopted the experimental lattice constants of bulk LaNiO$_{2}$\cite{crespin_jssc2005}.
The difference between L$_{\mathrm{large}}$ and L$_{\mathrm{exp}}$ yields $\Delta a$ $\approx$0.04$\textrm{\AA}$ and $\Delta c$ $\approx$0.1$\textrm{\AA}$ for in-plane and out-of-plane, respectively.
We define the smaller lattice L$_\mathrm{small}$ ($a$ = 3.88 $\textrm{\AA}$, $c$ = 3.18 $\textrm{\AA}$) to be [L$_{\mathrm{exp}}-(\Delta a, \Delta c)$].
We, therefore, examine the effect of lattice modulation on the Kondo screening mainly due to the expansion along the $c$-axis.
All lattice constants are shown in Fig.~\ref{Fig_2}(a). 
Note that the geometry optimization using many-body methods is currently not feasible.
Although our test with those lattice constants does not provide quantitative results comparable to the experimental data, we expect to uncover some distinct features induced by the variation of interlayer distance. 
The examination of the lattice modulation effect should help the understanding of the physics in the Sr-doped film with larger lattice. The measured data in Nd$_{1-x}$Sr$_{x}$NiO$_{2}$/SrTiO$_{3}$\cite{danfeng_prl2020} indicate that the $c$-axis lattice constant increased from 3.28 $\textrm{\AA}$ to 3.42 $\textrm{\AA}$ by Sr-doping $x$ $=$ 0.25. For Nd$_{0.8}$Sr$_{0.2}$NiO$_{2}$/SrTiO$_{3}$, the $c$-axis lattice constant becomes smaller, as the film is thicker: from $\sim$3.42 $\textrm{\AA}$ for the 4.6-nm film to $\sim$3.36 $\textrm{\AA}$ for the 15.2-nm film\cite{zeng_ncomm2022}.

Figure~\ref{Fig_2}(b) shows that the spectral weight of Nd-4$f$ and the band kinks at the Fermi level are found to be significantly weaker in L$_\mathrm{large}$, which indicates the Kondo effect diminished by the increase of inter-atomic distance.
This tendency is also described by the local total angular momentum susceptibility $\chi_{\mathrm{loc}}^{J_Z}$ of Nd-4$f$ behavior in Fig.~\ref{Fig_2}(a). $\chi_{\mathrm{loc}}^{J_Z}$ is given as
\begin{equation}
  \begin{split}
\chi_{\mathrm{loc}}^{J_Z}=\int_0^\beta d\tau\langle J_z(\tau)J_z(0)\rangle.
  \end{split}
\end{equation}
The $\chi_{\mathrm{loc}}^{J_Z}$ of Nd-4$f$ in L$_\mathrm{small}$ and  L$_\mathrm{exp}$ deviate from the Curie-Weiss behavior at $\sim$400 K, which indicates the onset temperature of the Kondo scattering process\cite{byung_ute2}. On the contrary, in L$_\mathrm{large}$, $\chi_{\mathrm{loc}}^{J_Z}$ of Nd-4$f$ keeps dropping with lowering temperature, which indicates that Kondo screening effect disappears\cite{byung_ute2}.

The lattice modulation effect on Kondo screening can be also understood through the calculation of self-energy ($\Sigma$), as shown in Fig.~\ref{Fig_2}(c).
At 900 K, in all three lattices, the imaginary part (Im) of $\Sigma$ of Nd-$f_{\alpha}$ exhibit a singularity on the imaginary frequency axis, which indicates that the electronic structure at high temperature is governed by Mott-like physics.
It also leads to the peak of Im $\Sigma$ at Fermi-level on the real frequency axis and the appearance of a gap in the DOS.
On the other hand, at a lower temperature, the self-energy exhibits Fermi-liquid-like behavior giving rise to a formation of quasi-particle peak at the vicinity of the Fermi level.
The high-$T$ Mott-like feature becomes pronounced in L$_\mathrm{large}$ owing to the suppressed inter-orbital hopping, which is manifested by the reduced hybridization function, as shown in Supplementary Fig. S2.
The prominent Mott-like characteristic of Nd-4$f$ in L$_\mathrm{large}$ hinders the formation of quasi-particle peak at the Fermi level, weakening the Kondo screening.
Our results are reminiscent of Fermi-liquid behavior of 5$f_{5/2}$ states with larger Kondo scale than the 5$f_{7/2}$ states which are at the edge of a Mott transition in PuCoGa$_{5}$\cite{brito2018orbital}. Interestingly, the two features are competing in the same states of NdNiO$_{2}$.

Our results propose that the Kondo temperature driven from Nd-4$f$ decrease with the increase of inter-atomic distance, which gives an explanation to the experimental data reported for the electrical resistivity upturn at low temperature.
The upturn was measured around 70 K for NdNiO$_{2}$ film ($c$-axis lattice constant: 3.31 $\textrm{\AA}$)\cite{danfeng_nature2019}, whereas insulating resistivity behavior arises below 300 K in NdNiO$_{2}$ bulk with smaller $c$-axis lattice constant of 3.24 $\textrm{\AA}$\cite{li_commat2020}. 
While the Kondo effect is not the only possibility for the resistivity upturn, these observations could indicate higher onset temperature of Kondo scattering in NdNiO$_{2}$ bulk than film, and be tied to our result of weaker Kondo effect in larger lattice.

\begin{figure*}[ht]
\centering
\includegraphics[width=0.84 \textwidth]{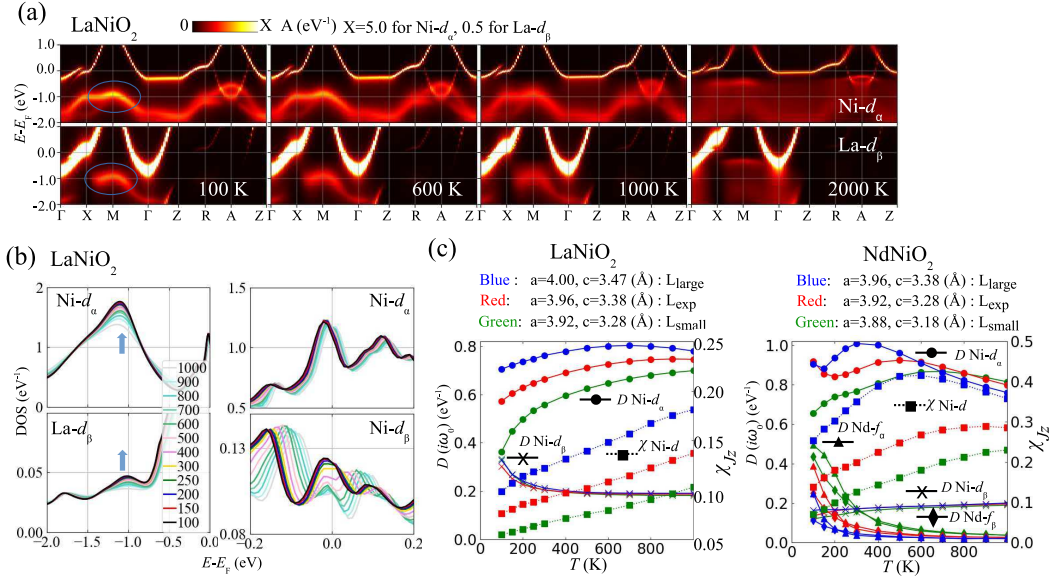}
\caption{\label{Fig_3}  \textbf{Temperature dependence of $d$-$d$ hybridization effect.} (a) Ni-$d_{\alpha}$ and La-$d_{\beta}$ projected spectral functions at 100, 600, 1000 and 2000 K. (b) Ni-$d_{\alpha}$, La-$d_{\beta}$, Ni-$d_{\alpha}$ and Ni-$d_{\beta}$ projected density of states. The temperature unit is K. (see blue arrows) As temperature decreases, the hybridized peaks of Ni-$d_{\alpha}$ and La-$d_{\beta}$ are increased, which makes the DOS of Ni-$d_{\alpha}$ at $E_{\textrm{F}}$ lower. (c) The DOS at $E_{\textrm{F}}$ $D(\mathrm{i}\omega_{0})$ (solid lines) of Ni-($d_{\alpha}$,$d_{\beta}$) and Nd-($f_{\alpha}$,$f_{\beta}$), and the local total angular momentum susceptibility
$\chi_{\mathrm{loc}}^{J_Z}$ (dashed lines) of Ni-3$d$ in three lattices of L$_\mathrm{large}$ (blue),
L$_\mathrm{exp}$ (red) and L$_\mathrm{small}$ (green), respectively, for LaNiO$_{2}$ and NdNiO$_{2}$. }
\end{figure*}

\textbf{$d$-$d$ hybridization in LaNiO$_{2}$ and NdNiO$_{2}$.}
LaNiO$_{2}$ has the same crystal structure as NdNiO$_{2}$, however La-4$f$ states are far away from the Fermi level. 
To investigate aforementioned hybridization effect, we first calculated the electronic structure of LaNiO$_{2}$. 
The two left panels of Fig.~\ref{Fig_3}(b) show DOS of Ni-$d_{\alpha}$ and La-$d_{\beta}$. In this work, we focus on the $d$-$d$ hybridization which manifests the evolution of Ni-$d_{\alpha}$ and La-$d_{\beta}$ peaks around -1.0 eV with decreasing the temperature. 
The peaks are absent at 2000 K, and developed 
along the X-M-R symmetry line with lowering temperature, as shown in Fig.~\ref{Fig_3}(a).
The apparent coincidence of peak positions of Ni-3$d$ and La-5$d$ is presented by one orbital projected DOS in Supplementary Fig. S3.
Due to formation of the hybridized peaks, Ni-$d_{\alpha}$ peak at the Fermi level shifted to lower energy with lowering temperature, as shown in the right top panel of Fig.~\ref{Fig_3}(b). 
The shift reduces the DOS of Ni-3$d$ (Ni-$d_{\alpha}$ $+$ Ni-$d_{\beta}$) at the Fermi level (see circle markers, red line in the left panel of Fig.~\ref{Fig_3}(c)),
resulting in decreased local total angular momentum susceptibility $\chi_{\mathrm{loc}}^{J_Z}$ of Ni-3$d$ (see square markers, dotted red line in the left panel of Fig.~\ref{Fig_3}(c)) with lowering temperature.
Here,
the DOS at the Fermi level was calculated by
\begin{equation}
  \begin{split}
  D(\mathrm{i}\omega_{0})=-\frac{1}{\uppi}\textrm{Im}G(\mathrm{i}\omega_{0}),
  \end{split}
\end{equation}
where $\omega_{0}$ is the first Matsubara frequency and $G$ is the calculated local Green's function. 

We tested the effect of lattice modulation on the $d$-$d$ hybridization.
We took three lattices L$_\mathrm{small}$, L$_\mathrm{exp}$, and L$_\mathrm{large}$, similarly to NdNiO$_{2}$. Here, L$_{small}$ uses the experimental lattice parameters of bulk NdNiO$_{2}$, L$_{exp}$ uses the experimental lattice constants of bulk LaNiO$_{2}$, and L$_{large}$ uses the L$_{exp}$ added with the difference between L$_\mathrm{exp}$ and L$_\mathrm{small}$ (see Fig.~\ref{Fig_3}(c)). 
As shown in Fig.~\ref{Fig_3}(c), the DOS at the Fermi level $D(\mathrm{i}\omega_{0})$ of Ni-$d_{\beta}$ are almost the same in magnitude, together with similar temperature dependence in all three lattices. This indicates a negligible effect of lattice modulation on Ni-$d_{\beta}$.
All $D(\mathrm{i}\omega_{0})$ of Ni-$d_{\beta}$ gradually increase upon cooling (see also the right bottom panel of Fig. 3(b)).
This behavior is reminiscent of evolution of quasi-particle peaks in strongly correlated materials, where the spectral weight at the Fermi level is transferred from the upper and lower Hubbard bands~\cite{byung_ute2,deng2019signatures,choi2013observation}.

On the contrary, as shown in Fig.~\ref{Fig_3}(c), all $D(\mathrm{i}\omega_{0})$ of Ni-$d_{\alpha}$ exhibit overall enhancement as the lattice is larger, indicating the significant lattice modulation effect on the $d$-$d$ hybridization, unlike Ni-$d_{\beta}$.
They increase upon cooling to a certain temperature, below which they start decreasing (Data calculated above 1000 K are not shown here. But this is clearly shown in NdNiO$_{2}$).
The temperature of maximum $D(\mathrm{i}\omega_{0})$ is proportional to the strength of the $d$-$d$ hybridization.
As the lattice is larger, L$_\mathrm{small}$ $\rightarrow$ L$_\mathrm{large}$, in LaNiO$_{2}$, the temperature of maximum $D(\mathrm{i}\omega_{0})$ of Ni-$d_{\alpha}$ decreases from above 1000 K to around 700 K, and
$D(\mathrm{i}\omega_{0})$ of Ni-$d_{\alpha}$ and $\chi_{\mathrm{loc}}^{J_Z}$ of Ni-3$d$ become higher. 
The hybridized peaks of Ni-$d_{\alpha}$ and La-$d_{\beta}$ are together weakened and shifted to higher energy (see Supplementary Fig. S3 and S4).
These indicate that as the lattice is larger, the $d$-$d$ hybridization between Ni-$d_{\alpha}$ and La-$d_{\beta}$ is weakened, making the DOS of Ni-$d_{\alpha}$ at the Fermi level higher.

The $d$-$d$ hybridization in NdNiO$_{2}$ is manifested in the same fashion as in LaNiO$_{2}$.
As lattice sizes are larger  L$_\mathrm{small}$ $\rightarrow$ L$_\mathrm{large}$, the $d$-$d$ hybridization peaks of Ni-$d_{\alpha}$ and Nd-$d_{\beta}$ become weaker. 
As shown in Fig.3(c),
$D(\mathrm{i}\omega_{0})$ of Ni-$d_{\alpha}$, and
$\chi_{\mathrm{loc}}^{J_Z}$ of Ni-3$d$ for both
LaNiO$_{2}$ and NdNiO$_{2}$ show the similar trend in response to the lattice modulations down to about 200 K, below which 
the $D(\mathrm{i}\omega_{0})$ of Ni-3$d$ and Nd-4$f$ increases due to 
the emerging Kondo effect (see Fig.~\ref{Fig_3} (c)). 
However, the temperatures of maximum $D(\mathrm{i}\omega_{0})$ of Ni-$d_{\alpha}$ and $\chi_{\mathrm{loc}}^{J_Z}$ of Ni-3$d$ in NdNiO$_{2}$ are much lower, about half, than those in LaNiO$_{2}$. 
As L$_\mathrm{small}$ $\rightarrow$ L$_\mathrm{large}$ in NdNiO$_{2}$, the temperature of maximum $D(\mathrm{i}\omega_{0})$ drops significantly from about 700 K to about 300 K.  
Thus the $d$-$d$ hybridization in NdNiO$_{2}$ is suggested to be much weaker than that in LaNiO$_{2}$.
This is also described by the weaker hybridization function in NdNiO$_{2}$, as shown in Supplementary Fig. S7.

\begin{figure*}[ht]
\centering
\includegraphics[width=0.5 \textwidth]{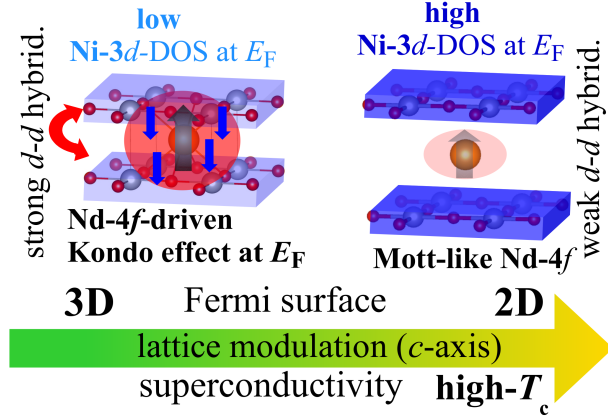}
\caption{\label{Fig_4} 
\textbf{Lattice modulation effect.} Schematic diagram illustrates relationship among the lattice modulation, the electronic interactions between Ni and rare-earth, and the 3D-to-2D transition of Fermi surface nature in NdNiO$_{2}$. 
Red curved arrow denotes the $d$-$d$ hybridization.
Black and blue arrows denote local magnetic moment of Nd-4$f$ and spins of conduction Ni-3$d$ electrons, respectively, whose strong-correlation induces Kondo cloud.
}
\end{figure*}

\textbf{Discussion} 

We have shown that the electronic structure at Fermi level in bulk nickelates has 3D nature through the Kondo effect driven by Nd-4$f$ and Ni-3$d$ at low temperature, {\it i.e.,} the Fermi surface of bulk NdNiO$_{2}$ consists of 3D multi-bands: i) Kondo band by Nd-4$f$, ii) Ni-$d_{\alpha}$, and iii) mixed bands of Nd-$d_{\beta}$, Ni-$d_{\beta}$ and Nd-$d_{\alpha}$.
The third one can be lifted from the Fermi level by hole doping\cite{byung_2020,botana_prx2020}.
According to these results and the experimental observation of insulating behaviors of non-superconducting Nd$_{0.8}$Sr$_{0.2}$NiO$_{2}$ bulk\cite{li_commat2020}, we suggest that the hole doped NdNiO$_{2}$ still has 3D multi-band by the Kondo bands of the Ni-$d_{\alpha}$ and Nd-4$f$ at the Fermi level. 
As schematically described by Fig.~\ref{Fig_4},
as the distance between Nd and Ni is larger, 
(i) the Kondo effect is suppressed or eliminated, thus
the 3-dimensional Kondo cloud around Fermi-liquid-like Nd-4$f$ can be vanished, by which the quasi-2D-like Fermi surface is formed,
(ii) the $d$-$d$ hybridization between Ni and rare-earth is weakened, by which the DOS of Ni-3$d$ at the Fermi level is increased.

The superconductivity of nickelates occurs only in films with the larger interlayer distance than that of bulk. In addition, the recent measure of significant angle-dependent $T_{\textrm{c}}$, also, emphasizes a more important role of quasi-2D bands comprised of Ni-3$d_{x^{2}-y^{2}}$ for the pair formations in nickelates\cite{wenjie_arxiv2022}. 
A recent theoretical work also shows the presence of dominant pairing instability on the Ni-3$d_{x^{2}-y^{2}}$ 2D single band channel\cite{xianxin_prb2020}.
It is suggested that quasi-2D-like Fermi surface is essential for unconventional superconductivity in cuprates\cite{bollinger_sst2016}.
These observations, reflecting our finding about lattice modulation effect, suggest that the 2D framework, applied to the cuprates, may also be valid for the nickelates, and the superconductivity in the nickelates can be promoted by the 3D-to-2D transition.

Therefore, the concerted effect of the 3D-to-2D transition and the increased DOS achieved by lattice modulation can boost superconductivity. These can give explanations to the following three experimental observations of appearance (absence) of superconductivity for the Nd$_{0.8}$Sr$_{0.2}$NiO$_{2}$ film (bulk) with the large (small) $c$-axis lattice constant.
We assume that these films and bulk have the same doping concentration, and only difference is in lattice size.
First, the Sr doping arise the lattice modulation effect of the monotonic increment of interlayer distance from 3.28 $\textrm{\AA}$ to 3.42 $\textrm{\AA}$ upon zero to 25$\%$ Sr doping in Nd$_{1-x}$Sr$_{x}$NiO$_{2}$/SrTiO$_{3}$ thin films, where superconducting dome appear for 0.125 $<$ $x$ $<$ 0.25\cite{danfeng_prl2020}. The interlayer distance is $\sim$3.38 $\textrm{\AA}$ for Nd$_{0.8}$Sr$_{0.2}$NiO$_{2}$/SrTiO$_{3}$.
Second, the superconductivity is observed in films with thickness ($c$-axis lattice constant) ranging from 4.6-nm ($\sim$3.42 $\textrm{\AA}$) to 15.2-nm ($\sim$3.36 $\textrm{\AA}$) in Nd$_{0.8}$Sr$_{0.2}$NiO$_{2}$/SrTiO$_{3}$\cite{zeng_ncomm2022}.
Third, no superconductivity is observed in Nd$_{0.8}$Sr$_{0.2}$NiO$_{2}$ bulk\cite{li_commat2020}. Its interlayer distance is 3.33 $\textrm{\AA}$, which is smaller than that of the films.

Our work sheds light on the superconductivity mechanism in nickelates and 
suggest that the superconductivity can be readily enhanced by engineering the 3D-to-2D transition through lattice modulation.
It provides with an immediate route to the manipulation of the superconductivity.

\section*{Methods}
\subsection{LQSGW and DMFT calculations\\}
We use $ab$-$initio$ linearized quasi-particle self-consistent GW (LQSGW) and dynamical mean field theory (DMFT) method~\cite{tomczak2015qsgw,choi2016first,choi2019comdmft} to calculated the electronic structure of LaNiO$_{2}$ and NdNiO$_{2}$ which crystallizes into tetragonal space group P4/mmm (No. 123)~\cite{crespin_jssc2005,hayward_sss2003}.  
The LQSGW+DMFT is designed as a simplified version of the full GW+DMFT approach~\cite{sun2002extended,biermann2003first,nilsson2017multitier}. 
It calculates electronic structure by using LQSGW approaches~\cite{kutepov2012electronic,kutepov2017linearized}. 
Then, it corrects the local part of GW self-energy within DMFT~\cite{georges1996dynamical,metzner1989correlated,georges1992hubbard}. 
We adopt experimental lattice constants of $a=$ 3.96 and $c=$ 3.38 $\textrm{\AA}$~\cite{crespin_jssc2005} and $a=$ 3.92 and $c=$ 3.28 $\textrm{\AA}$~\cite{hayward_sss2003} for LaNiO$_{2}$ and NdNiO$_{2}$, respectively. 
Using the two lattices, we generated artificially the smaller and the larger lattices for each nickelate to model lattice modulation. 
Other than the lattice parameters, we explicitly calculate all quantities such as frequency-dependent Coulomb interaction tensor and double-counting energy. 
Then, the local self-energies for Ni-3$d$ and Nd-4$f$ are obtained by solving two different single impurity models. 
La-5$d$ and Nd-5$d$ are treated within GW approximation. 
Test simulations, which treat rare-earth-5$d$ as strongly correlated orbitals within DMFT, show qualitatively similar Kondo effect (kink-like band structure at the Fermi level along $\Gamma$-Z) and interlayer hybridization effect (data are not presented) with presented results.
Spin-orbital coupling is included for all calculations.
For the details of method, please see the supplementary.

\begin{addendum}
 \item B.K. thanks Sangkook Choi for insightful discussion. This research used resources of the National Energy Research Scientific Computing Center (NERSC), a U.S. Department of Energy Office of Science User Facility operated under Contract No. DE-SC0021970.
 \item[Competing Interests] The authors declare no competing interests.
 \item[Correspondence] Byungkyun Kang~(email: byungkyun.kang@unlv.edu)

 \item[Author contributions] B.K. designed the project. B.K., H.K., Q.Z and C.P. wrote the manuscript. B.K. performed the calculations and conducted the data analysis. All authors discussed the results and commented on the paper.
\end{addendum}

\section*{REFERENCES}

\bibliography{ref}

\end{document}